\documentclass[12pt,preprint]{aastex}
\usepackage{amsmath}

\shorttitle{Studying MHD Turbulence with Synchrotron Polarization}

\shortauthors{ZHANG, LAZARIAN, LEE \& CHO}

\begin{document}

\title{Studying Magnetohydrodynamic Turbulence with Synchrotron Polarization Dispersion}

\author{Jian-Fu Zhang\altaffilmark{1,2,3}, Alex Lazarian\altaffilmark{2}, Hyeseung Lee\altaffilmark{2,4} \& Jungyeon Cho\altaffilmark{2,4}}
\altaffiltext{1}{Department of Astronomy and Institute of Theoretical Physics and Astrophysics, Xiamen University, Xiamen, Fujian 361005, China}\email{jianfuzhang.yn@gmail.com(JFZ); alazarian@facstaff.wisc.edu(AL); hsleewill@gmail.com(HL); jcho@cnu.ac.kr(JC)}
\altaffiltext{2}{Astronomy Department, University of Wisconsin, Madison, WI 53711, USA}
\altaffiltext{3}{Department of Physics, Tongren University, Tongren, Guizhou 554300, China}
\altaffiltext{4}{Department of Astronomy and Space Science, Chungnam National University, Daejeon, Republic of Korea}

\begin{abstract}
We test a new technique of studying magnetohydrodynamic (MHD) turbulence suggested
by Lazarian \& Pogosyan, using synthetic synchrotron polarization observations. This paper focuses on a one-point statistics, which is termed the polarization frequency analysis, that is characterized by the variance of polarized emission as a function of the square of wavelengths along a single line of sight. We adopt a ratio $\eta$ of the standard deviation of the line-of-sight turbulent magnetic field to the line-of-sight mean magnetic field to depict the level of turbulence. When this ratio is either large ($\eta\gg1$), which characterizes a turbulent field dominated region, or small ($\eta\lesssim0.2$), which characterizes a mean field dominated region, we obtain the polarization variance $\left<P^2\right>\propto\lambda^{-2}$ and $\left<P^2\right>\propto\lambda^{-2-2m}$, respectively. At small $\eta$, i.e., the mean field dominated region, we successfully recover the turbulent spectral index by the polarization variance. We find that our simulations agree well with the theoretical prediction of Lazarian \& Pogosyan. With existing and upcoming data cubes from the Low-Frequency Array for Radio astronomy (LOFAR) and Square Kilometer Array (SKA), this new technique can be applied to study the magnetic turbulence in the Milky Way and other galaxies.

\end{abstract}

\section{Introduction}
\label{intro}

Magnetohydrodynamic (MHD) turbulence is ubiquitous in astrophysical environments (evidence from electron densities: \citealt{Armstrong95,ChepL10,BL12}, from spectral lines: \citealt{Larson81,LP00,L09,ChepL10,ChepL15} and from synchrotron fluctuations: \citealt{Cho10,Gaensler11,Burkhart12,Iacobelli14}, see also \citealt{Elmegreen04,Mckee07} for reviews). It plays a critical role for key astrophysical  processes, such as star formation, acceleration and propagation of cosmic rays, heat transport, and magnetic reconnection (see \citealt{ChoLV03,Mac04,Laz15}). A significant progress has been achieved in understanding the theory of MHD turbulence (e.g., \citealt{Goldreich95,Cho03}, see also \citealt{BranL13,BereL15} for recent reviews), and its implications, in particular, the physics of turbulent reconnection (e.g., \citealt{LV99,Kowal09,Laz15} for a recent review). However, the theory of MHD turbulence is a developing field with many outstanding  questions. For instance, even the spectral slope of the Alfv\'{e}nic turbulence is  a subject of strong debates (e.g., see \citealt{Bere14} and references therein). The changes of the spectral slope with the physical conditions in plasmas are mostly unknown quantities that is difficult to study with the present time numerical resolutions.
Studies of those  using observations are really advantageous.

The seminal theoretical work of \cite{Goldreich95} predicted that the spectrum of MHD turbulence follows the Kolmogorov spectrum with $k^{-11/3}$ in terms of 3-dimensional (3D) spectrum. However, the spectrum of MHD turbulence may become different in some astrophysical environments. For example, the magnetic spectrum in terms of 3D spectrum is claimed to be $k^{-3}$ in the case of viscosity-damped turbulence when the viscosity to resistivity ratio is much larger than 1 (\citealt{ChoLV02,ChoLV03,Laz04}). The recovery of the spectral index of magnetic turbulence is the focus of this paper. In addition, spectra of both density and velocity are also important to reveal information on MHD turbulence. The former can be obtained directly from the column density maps (e.g., \citealt{Stutzki99}). However, the study of the latter is relatively complex. The new techniques to study the velocity spectrum, which are termed velocity channel analysis (VCA) and velocity coordinate spectrum (VCS), have been proposed (\citealt{LP00,LP04,LP06,LP08}), and successfully testified and applied to observations (\citealt{Laz01,Padoan06,Padoan09,ChepL09,Chep10,ChepL15}).

The overall spectral shape and spectral index of MHD turbulence can provide very valuable information, such as the sources and sinks of turbulence, the cascading processes of energy. Due to complexity and uncertainty of MHD turbulence, it is important to have techniques to study turbulence from observations. It is no doubt that the correct understanding of MHD turbulence can help us to shed light on some intrinsic properties of astrophysical processes, and even challenge paradigms of some traditional models,
e.g.~evolution of molecular clouds (see \citealt{Mest65,Mous76,Stone98,Laza14}).

When the non-thermal relativistic electrons propagate in turbulent magnetic fields they emit polarized synchrotron radiation, which carry information on the statistics of turbulent flows. Therefore, synchrotron fluctuations are promising tool for investigating MHD turbulence. A theoretical description of the fluctuations of synchrotron intensity  arising from magnetic turbulence is provided in \cite{LP12} (henceforth LP12) and it was tested  \cite{Herron16} by using synthetic observations.

The study on synchrotron polarization fluctuations together with Faraday rotation measure (RM) has been initiated in \cite{Burn66}, and from then on many studies are dedicated to this subject (e.g., \citealt{Brentjens05,Frick10,Frick11,Beck12}, see \cite{Heald15} for a recent review).

Very recently, a novel technique of the statistic description of synchrotron polarization fluctuations is suggested to investigate spectrum and anisotropy of MHD turbulence (\citealt{LP16}, henceforth LP16). This new technique considers the complex polarized intensity $P$ ($\equiv Q+iU$, where $Q$ and $U$ are Stokes parameters) as a function of the intrinsic polarized intensity density $P_{\rm i}$, and then integrates along the line of sight in physical space\footnote{In addition, the Faraday dispersion function is defined as a Fourier transform of the polarization surface brightness with regard to $\lambda^2$ variable in Appendix D of LP16, which is used to study MHD turbulence  by the Fourier transform from the wavelength $\lambda$ space to the physical space.},
\begin{equation}
P(\textbf{\textit{X}},\lambda^2)=\int^{L}_{0}{\rm d}z P_{\rm i}(\textbf{\textit{X}},z)e^{2i\lambda^2\Phi(\textbf{\textit{X}},z)}, \label{PI}
\end{equation}
where $P(\textbf{\textit{X}},\lambda^2)$ is a function of the 2D spatial separation, $\textbf{\textit{X}}$, of the direction of measurements and the square of wavelength $\lambda^2$, and $L$ is the extent of the turbulent source along the line of sight. $\Phi(\textbf{\textit{X}},z)\propto\int_{0}^{z}n_{\rm th}(z')B_{\rm \|}(z'){\rm d}z'$ is called the Faraday RM or Faraday depth, where $n_{\rm th}$ is the density of thermal electrons and $B_{\rm \|}$ is the component of the magnetic field along the line of sight direction. Based on an analytical method of statistical descriptions, they proposed several versions of the technique that can be employed to obtain spectral slopes and correlation scales of the underlying MHD turbulence, and the spectrum of Faraday rotation fluctuation. Two main versions of the technique are polarization frequency analysis (PFA), which makes use of the change of the variance of polarization intensity (or its derivative) as a function of the square of the wavelength $\lambda^2$, and polarization spatial analysis (PSA), which makes use of the spatial correlations of the polarization intensity (or its derivative) at the same wavelength as a function of the spatial separation $\textbf{\textit{X}}$. LP16 stresses that the two techniques have some analogies with their VCS and the VCA techniques for turbulence studies using spectroscopic data. These techniques have been successfully applied to study turbulence with atomic hydrogen and CO data.

The purpose of this paper is to test the theoretical work of LP16 in order to open ways for applying this new technique to the studies of MHD turbulence in the Milky Way, distant galaxies and even clusters of galaxies. This work is timely because upcoming telescopes, such as the Low-Frequency Array for Radio astronomy (LOFAR) and Square Kilometer Array (SKA) operating in a wide wavelength range, can be used to study MHD turbulence; the origin and evolution of cosmic magnetism is a key science project of SKA (\citealt{Beck15c}). In this paper, we focus on testing the PFA theory; the test for PSA is carried out in Lee et al. (2016).

Next section is devoted to methods of statistical descriptions for studying MHD turbulence, and a brief description of numerical techniques is provided in Section 3. We present our results in Section 4, discuss the results of statistical test in Section 5, and summarize our findings in Section 6.

\section{Statistical Description of Turbulence}

MHD turbulence is an extremely complex chaotic non-linear process and exhibits diverse properties on a microscopic level, but it allows a macroscopic treatment in a simple statistical description (\citealt{Biskamp03}), which can reveal regular features behind chaotic phenomena of turbulent magnetic fluctuations. In fact, the method of statistical description of MHD turbulence is widely used. It is claimed as an adequate and concise way to shed light on important properties of interstellar turbulence (LP16).

Practically, structure and correlation functions of (any) physical variable $P(\textbf{\textit{X}})$ have been traditionally employed for studying MHD turbulence. The correlation function is given by
\begin{equation}
{\rm CF}(\textbf{\textit{R}})=\left< P(\textbf{\textit{X}}_{\rm 1})P(\textbf{\textit{X}}_{\rm 2})\right>, \label{CFp}
\end{equation}
where, $\left <...\right>$ indicates an average over the entire volume of interest. As for the structure function, there is a wider variety one can measure. In the case of an isotropic MHD turbulence, a second-order scenario is commonly used, and written as
\begin{equation}
{\rm SF}(\textbf{\textit{R}})=\left< (P(\textbf{\textit{X}}_{\rm 1})-P(\textbf{\textit{X}}_{\rm 2}))^2\right>=2\left [{\rm CF}(0)-{\rm CF}(\textbf{\textit{R}})\right ]. \label{SFqe}
\end{equation}
As shown in Equation (\ref{SFqe}), the second-order structure function is formally related to the correlation function.
On the other hard, the power spectrum, $E(k)$, is obtained by a Fourier transform of $P(\textbf{\textit{X}})$ from the physical space to the wavenumber space, and given by
\begin{equation}
E(k)=\frac{1}{2}P(k)P^*(k),
\end{equation}
where the symbol `*' denotes the complex conjugate of $P(k)$.
Generally, from a spatial point of view, 
the measures in this paragraph belong to two-point statistics.

Just like that stressed in Section \ref{intro}, we intend to test the PFA theory in LP16, which is essentially spectroscopic, requiring sufficient frequency resolution and coverage. This technique is to measure the correlation of polarization at different wavelengths along a fixed line of sight\footnote{
       Nevertheless, we will need averaging over many lines of sight to reduce statistical
       noise (see Equation (\ref{var1})).}.
In other words, this is a variance of polarization intensity, that is, a one-point statistics and a special case of the correlation function at $\textbf{\textit{R}}=0$. We now describe briefly the main theoretical results of LP16 related to the current work as follows.

On the basis of Equation (\ref{PI}), the variance of polarization as a function of wavelength is written as
\begin{equation}
\begin{aligned}
\left < P(\lambda^2)P^*(\lambda^2)\right>=\left<P^2(\lambda^2)\right>=\int_{0}^{L} {\rm d}z_{\rm 1}\int_{0}^{L}{\rm d}z_{\rm 2} \times\\
e^{2i\bar{\phi}\lambda^2(z_{\rm 1}-z_{\rm 2})} \left < P_{\rm i}(z_{\rm 1})P^*_{\rm i}(z_{\rm 2})e^{2i\lambda^2[\Phi (z_{\rm 1})-\Phi (z_{\rm 2})]}\right>, \label{var1}
\end{aligned}
\end{equation}
where
$\bar{\phi}=\left<n_{\rm th}(z)B_{\rm \parallel}(z)\right>$ is an average of the RM density. After considering two assumptions, that is, one is $\phi$ to be a Gaussian quantity, and the other is to neglect the correlation between the intrinsic polarization fluctuation $P_{\rm i}$ and the Faraday RM, Equation (\ref{var1}) can further factorized into
\begin{equation}
\begin{aligned}
\left<P^2(\lambda^2)\right>=\int_{0}^{L} {\rm d}z_{\rm 1}\int_{0}^{L}{\rm d}z_{\rm 2}e^{2i\bar{\phi}\lambda^2(z_{\rm 1}-z_{\rm 2})} \times\\
\left < P_{\rm i}(z_{\rm 1})P^*_{\rm i}(z_{\rm 2})\right> e^{-4\lambda^4D_{\rm \Delta \Phi}(0,z_{\rm 1}-z_{\rm 2})}, \label{var2}
\end{aligned}
\end{equation}
where , $\left< P^2 \right> \equiv \left< PP^* \right>$ and
$D_{\rm \Delta \Phi}$ is the structure function of Faraday RM and was intensively studied in Section 3.2 of LP16. LP16 investigated different regimes of magnetic turbulence by changing the ratio $\sigma_{\phi}/\bar{\phi}$, which can appropriately characterize the transition of magnetic turbulence from the case when RM is dominantly random to the case uniform; here, $\sigma_{\phi}$ is the root mean square of Faraday RM density fluctuations.

In the case of $\bar{\phi}>\sigma_{\phi}$, the variance of polarization intensity is subjected to
\begin{equation}
\left<P^2(\lambda^2)\right> \propto \lambda^{-2-2m} \label{varslope1}
\end{equation}
at long wavelength regime, where $m$ is the characteristic scaling slope of magnetic fluctuations (for instance $m=2/3$ would correspond to Kolmogorov scaling). Obviously, Equation (\ref{varslope1}) reveals asymptotically the slope of the transverse magnetic field. In the opposite case, $\bar{\phi}<\sigma_{\phi}$, corresponding to the long wavelength regime,
\begin{equation}
\left<P^2(\lambda^2)\right> \propto \lambda^{-2} \label{varslope2}
\end{equation}
follows the universal law $\lambda^{-2}$ of Faraday rotation scaling. The weighted derivative of Equation (\ref{varslope2}) would reflect the scaling slope of the transverse magnetic field. As described however in LP16, it is not easy to recover properties of magnetic fluctuations from observational data cubes
due to the effects of noise in the data.

Testing the above expressions is the main goal of our study. In addition, we also present statistical studies for both Faraday depolarization and the variance of the polarization derivative as a function of wavelength. These studies would reveal the
statistics of magnetic fluctuations  from synchrotron polarization and Faraday rotation.

\section{Numerical Technique}
\label{NumTech}
\subsection{Synthetic Data Generation}
The technique we use to generate 3D synthetic data cubes for magnetic fields and the density of thermal electrons is presented here. We generate data ($\widetilde{\textbf{B}}(\textbf{k})$'s) in the wavenumber space and then transform them into the physical space. The magnetic field is assumed as
\begin{equation}
\textbf{\textit{B}}(\textbf{\textit{r}})=\sum_{k_{\rm min}\leq |\textbf{\textit{k}}|\leq k_{\rm max}}\widetilde{\textbf{\textit{B}}}(\textit{\textbf{k}}){ e}^{i\textbf{\textit{k}}\cdot \textit{\textbf{r}}}
\end{equation}
with the condition of the solenoidal vector field $\triangledown\cdot \textit{\textbf{B}}=0$, where $\textbf{\textit{k}}$ is the wave vector and $\textbf{\textit{r}}$ is the position vector.
The Fourier coefficient $\widetilde{\textbf{\textit{B}}}(\textit{\textbf{k}})$ is defined as $\widetilde{B}_{n}(\textit{\textbf{k}})=|\widetilde{B}_{n}(\textit{\textbf{k}})|{ e}^{i\zeta}$, where, the subscript $n=1,2,3$ indicates three directions of Cartesian coordinate. The amplitude follows a power-law: $|\widetilde{\textbf{\textit{B}}}(\textit{\textbf{k}})|\propto k^{-\beta}$.
The phase factor $\zeta$ is randomly distributed.
The generation of the data cube for the thermal electron density is similar but slightly easier because of no constraint for divergence-free condition.

Generating 3D data cubes requires large computational resources, which makes
it difficult to achieve high numerical resolution.
Hence, we also create both 1D and 2D synthetic data with  higher resolutions.
The procedure for generating the 1D and 2D data is similar to the one for 3D data.
That is, we generate Fourier coefficients of the form
  $\textbf{F}(\textbf{k})=\textbf{A}(\textbf{k})[{\rm cos}\varpi+i{\rm sin}\varpi]$
 in the wave-vector  space and
carry out Fourier transform from the $k$-space to the physical space.
Here $\textbf{k}$ is either a 2D vector or a 1D scalar,
$\varpi=2\pi\tau$, and $\tau$ is a random number between 0 and 1.

\subsection{Radiation Aspects}  \label{sect:3.2}
We give here a brief expression on numerical aspects for radiative processes. The distribution of the non-thermal relativistic electrons is assumed to be $N(\varepsilon)\propto \varepsilon^{-p}$, where $p$ is the spectral index of the electrons and $\varepsilon$ is their energy. The formulae of synchrotron radiation in Appendix A of LP16 are adopted to calculate Stokes parameter $I$, $Q$ and $U$. As seen in these formulae,
the Stokes parameters are proportional to
   $ |B_{\rm \perp}|^\gamma $,
   where $B_{\rm \perp}$ is the perpendicular component of the magnetic field
   and the exponent $\gamma$ is related to the electron index of the electron distribution
   $p$ by the relation $\gamma=(p+1)/2$. We stress that we are using the original formula of polarization intensity to test the theoretical work of LP16. In our work, related physical quantities are calculated in arbitrary units because they show only magnitude differences and do not change our statistical investigation results.

\subsection{Observational Effects}
To simulate realistic observations, we also consider the effects of a finite telescopic
angular resolution and
a random noise.
We convolve
the original polarization intensity (i.e., Stokes $Q$ and $U$) `maps' with a Gaussian kernel in order to study the influence of finite angular resolution.
For the noise, we generate a map of Gaussian noise and add it to
the original polarization intensity map.
We adjust the level of the noise  by considering the signal to noise ratio.
The resulting polarization maps (with noise) are  smoothed with
a Gaussian beam, whose
full-width at half maximum (FWHM) is equal to 3 pixels.
We fix this resolution because usual surveys have a typical value of 3 pixels corresponding to the FWHM of the peak of telescope beam, resulting in a standard derivation of $\sigma=3/2.35\simeq1.3$. To determine the signal to noise ratio of the final map, we also convolve the original polarization map with the same Gaussian beam (i.e., $\sigma \simeq1.3$) and then subtract it from the final map.

\begin{figure*}[]
    \begin{center}
        \begin{tabular}{ccc}
            \hspace{-0.79cm}
            \includegraphics[angle=90,width=50mm,height=50mm]{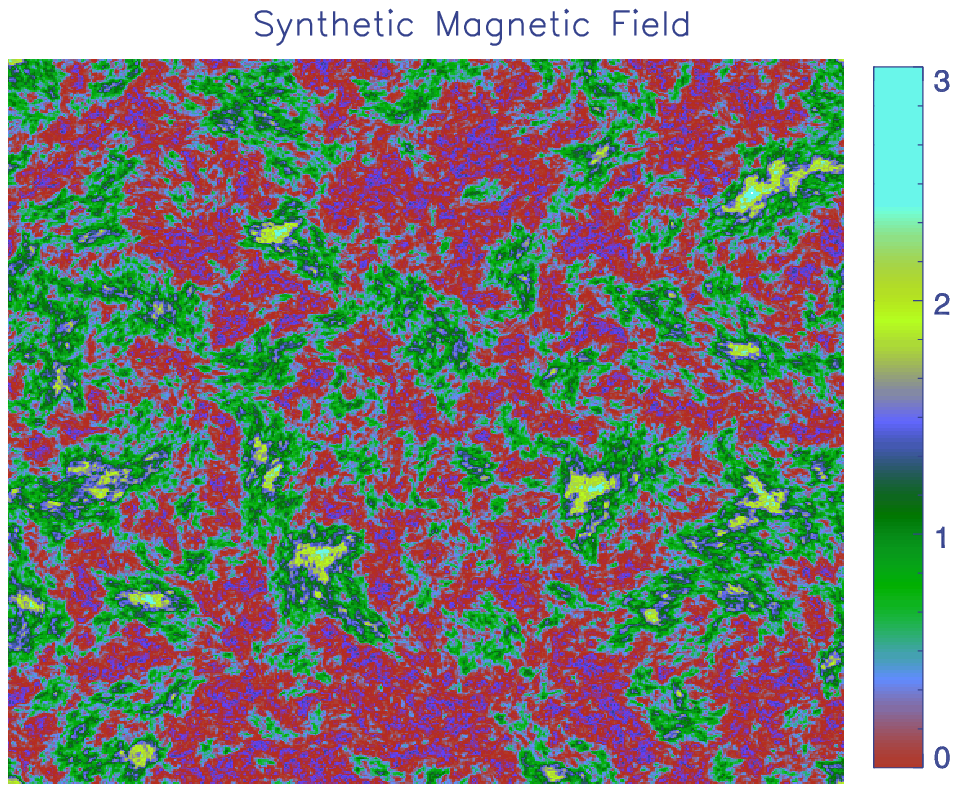}& \ \ \ \ \ \ \
            \hspace{-0.79cm}
            \includegraphics[angle=90,width=50mm,height=50mm]{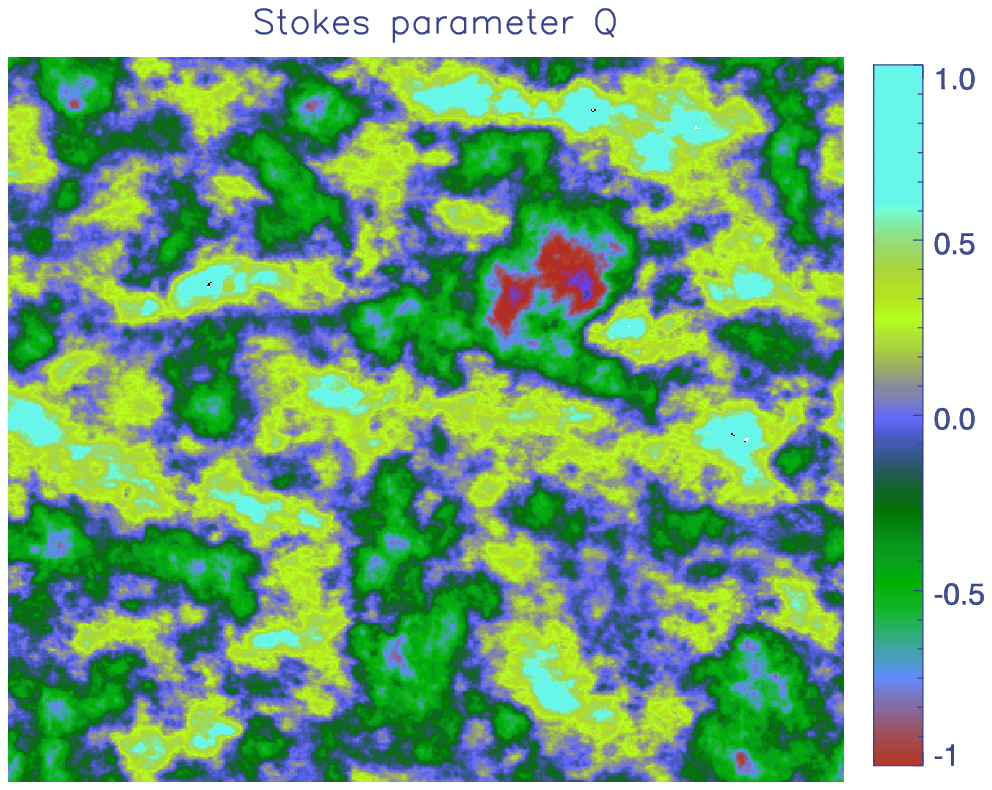}& \ \ \ \ \ \ \
            \hspace{-0.79cm}
            \includegraphics[angle=90,width=50mm,height=50mm]{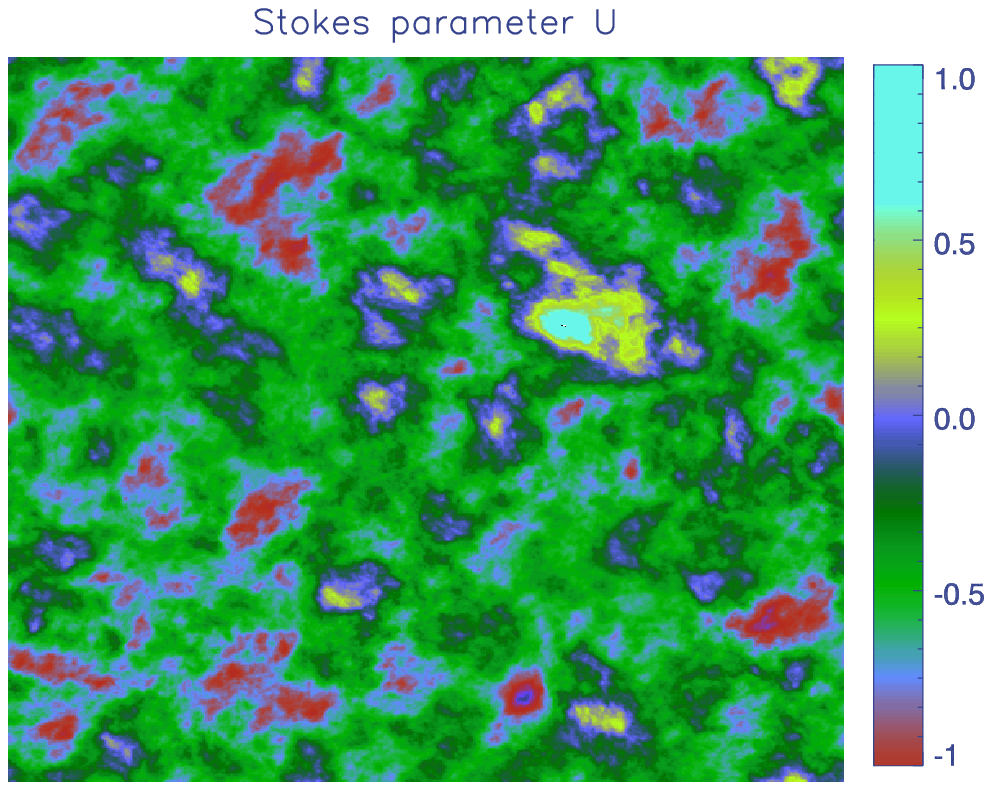}
        \end{tabular}
    \end{center}
    \caption{Left panel presents a cross-section of the synthetic magnetic field $B_{\perp}$ that has 512 pixels along each side and zero mean magnetic field. The maps of Stokes parameter $Q$ (middle panel) and $U$ (right panel) are obtained by invoking the magnetic field  structures in the left panel and integrating spatial separation along the $z$-axis. The spectral index of relativistic electrons is $p$=3, which links to $B_{\perp}^\gamma$ by $\gamma=(p+1)/2$. The units of the magnetic field strength, and of the Stokes parameter $Q$ and $U$ are arbitrary.}  \label{figs:MQU}
\end{figure*}

\section{Results}
\label{results}
We first provide maps of Stokes parameters from a 3D data cube in Section \ref{basic}. And then the results of one-point statistics are presented in detail in the following sections.

\subsection{Basics}
\label{basic}
We in this section use a 3D synthetic data cube with 512 pixels along each side and zero mean magnetic field $\left<B_z\right>=0$, which simulates an MHD turbulence flow with an isotropic Kolmogorov scaling, $k^{-11/3}$, to study statistics of polarization fluctuations. In Figure \ref{figs:MQU}, a map is plotted in the left panel to depict a cross-section of the perpendicular component of magnetic field, $B_{\rm \perp}=\sqrt{B_{x}^2+B_{y}^2}$. As shown on the map, the distribution of magnetic field is
homogeneous and isotropic. Using two 3D data cubes for $B_{x}$ and $B_{y}$, we obtain maps of Stokes parameters $Q$ (middle panel) and $U$ (right panel) by integrating along the $z$-axis. It should be noticed that the units of magnetic field and
the Stokes parameters, $Q$ and $U$, are arbitrary. The spectral index of relativistic electrons is fixed as $p$=3, which gives $B_{\perp}^{\gamma}=B_{\perp}^{2}$. We find that both $Q$ and $U$ have positive and negative values.
It seems that Stokes $Q$ has a larger probability for positive values but $U$ for negative values. In general, a positive $Q$ implies that the electric field vector is preferentially aligned with the  $x$-axis. Then a negative $Q$ means a preferential alignment with the $y$-axis. Stokes $U$ parameter has a similar meaning, just with a 45 degree rotated reference system.

We here used a zero mean magnetic field $\left<B_z\right>=0$, corresponding to a turbulent magnetic field dominated region, to carry out basic simulations. In the next sections, we will consider non-zero mean magnetic fields to perform related simulations. In this work, we use the ratio $\eta$ of a standard derivation $\sigma_z$ of the line-of-sight turbulent magnetic field to the line-of-sight mean magnetic field $\left<B_z\right>$ to characterize to the level of different magnetic turbulence. A large $\eta$ will indicate a turbulent magnetic field dominated region and a small $\eta$ will stand for a mean field dominated region.

\begin{figure} \begin{center}
        \includegraphics[scale=0.7,angle=90]{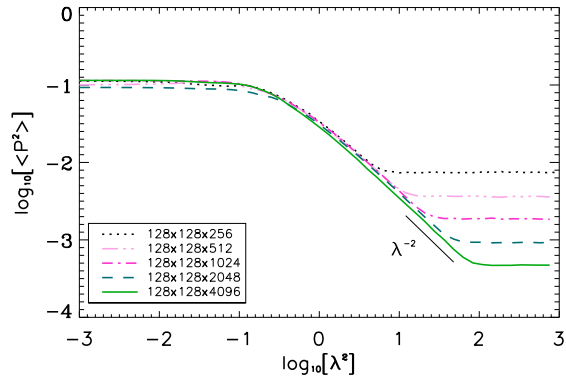}
\caption[ ]{Polarization variance as a function of the square of wavelength $\lambda^2$ in 3D case. The initial sets are $\beta=11/3$ and $\eta=\infty$.  } \label{fig:var3dres} \end{center}
\end{figure}

\begin{figure} \begin{center}
        \includegraphics[scale=0.7,angle=90]{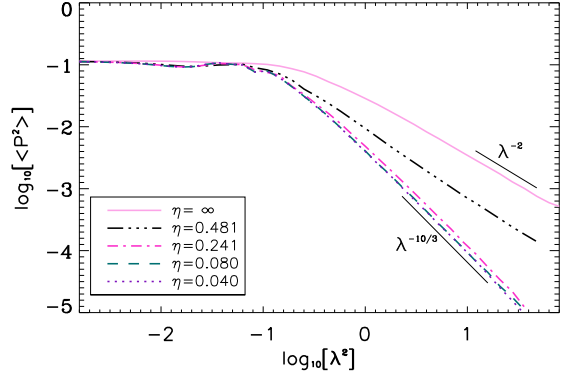}
\caption[ ]{The transition of polarization variance from the stochastic RM dominant regime for $\eta=\infty$, to the uniform RM dominant one for $\eta=0.080$ in 3D case. The initial set corresponds to Komolgorov scaling of $\beta=11/3$. The highest resolution 3D data cube is used.} \label{fig:var3dmean} \end{center}
\end{figure}

\begin{figure}
    \begin{center}
        \includegraphics[scale=0.7,angle=90]{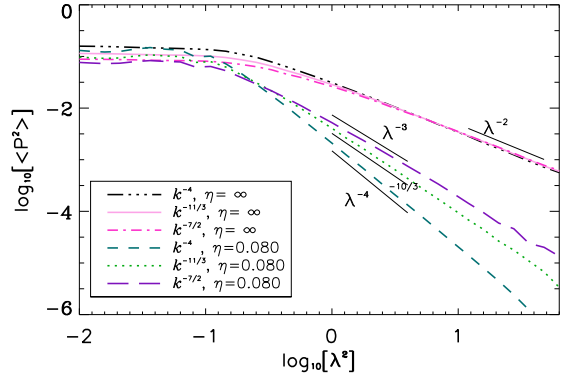}
\caption[ ]{Polarization variance for different MHD scaling indices, $\beta=4$, 11/3 and 7/2, as a function of the square of wavelength $\lambda^2$ in 3D case. $\eta=\infty$ and 0.080 characterize different regions of Faraday dispersion.
The highest resolution 3D data cube is used.
} \label{fig:var3dks}
    \end{center}
\end{figure}

\begin{figure} \begin{center}
        \includegraphics[scale=0.7,angle=90]{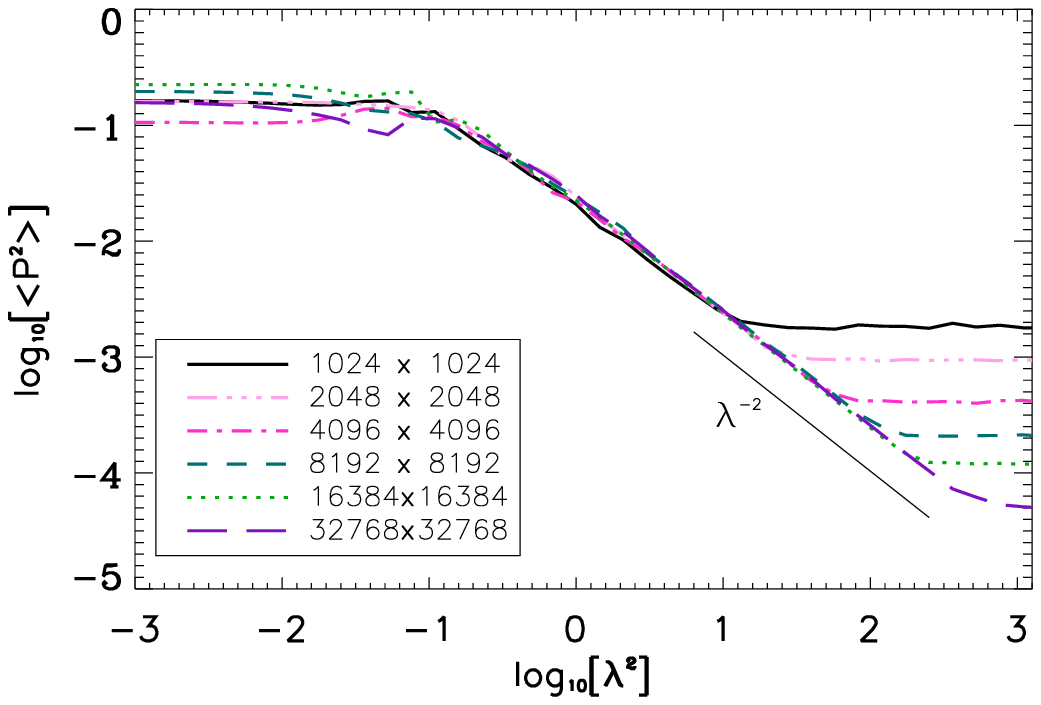}
\caption[ ]{Same as Figure \ref{fig:var3dres} but for 2D case with different resolutions. } \label{fig:var2dres} \end{center}
\end{figure}

\begin{figure}
    \begin{center}
        \includegraphics[scale=0.7,angle=90]{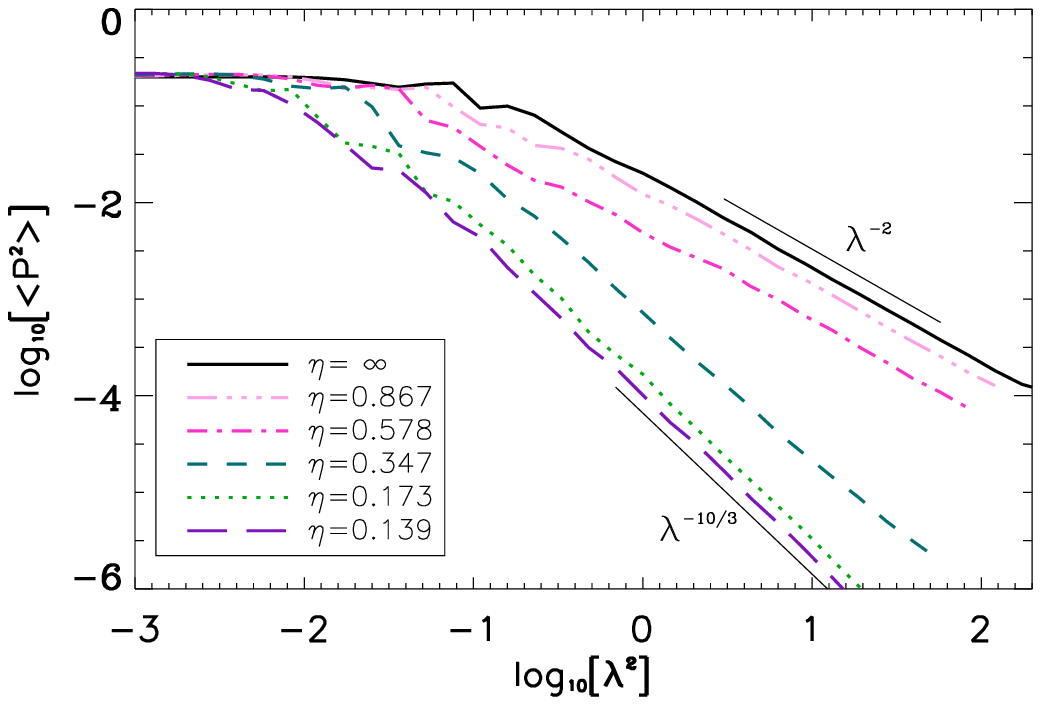}
\caption[ ]{Same as Figure \ref{fig:var3dmean} but for 2D case with different values of the mean magnetic field.
The highest resolution 2D data is used.} \label{fig:var2dmean}
    \end{center}
\end{figure}

\begin{figure}
    \begin{center}
        \includegraphics[scale=0.7,angle=90]{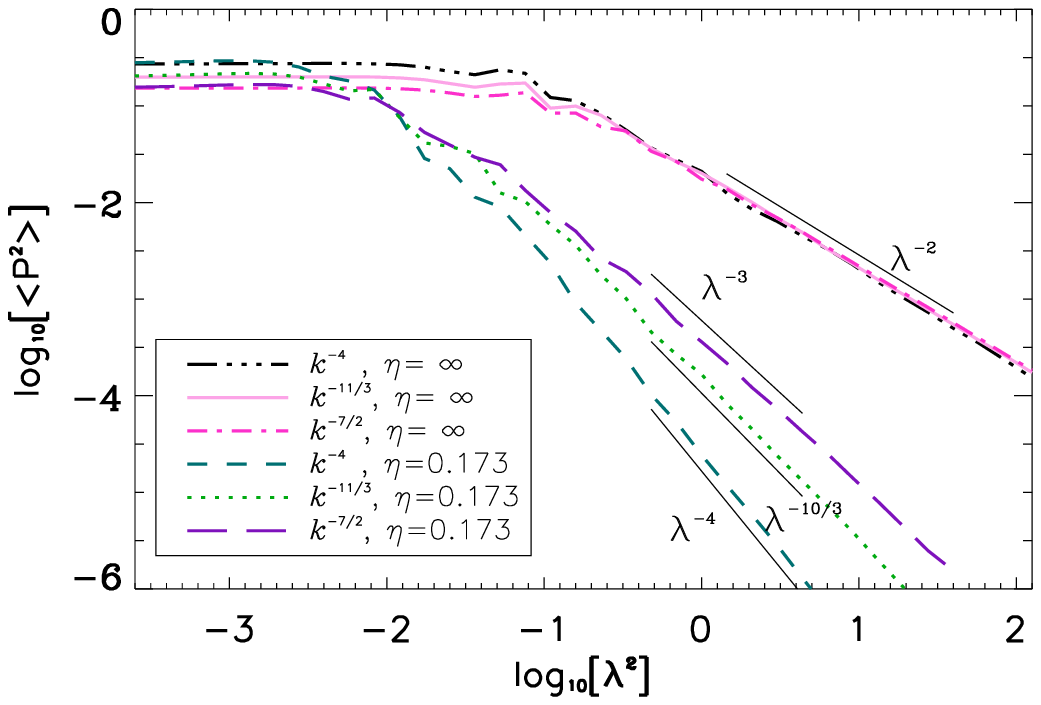}
\caption[ ]{Same as Figure \ref{fig:var3dks} but for 2D case. $\left<B_z\right>=0$ and 20 characterize different regions of Faraday dispersion.
The highest resolution 2D data is used.} \label{fig:var2dks}
    \end{center}
\end{figure}

\begin{figure} \begin{center}
        \includegraphics[scale=0.7,angle=90]{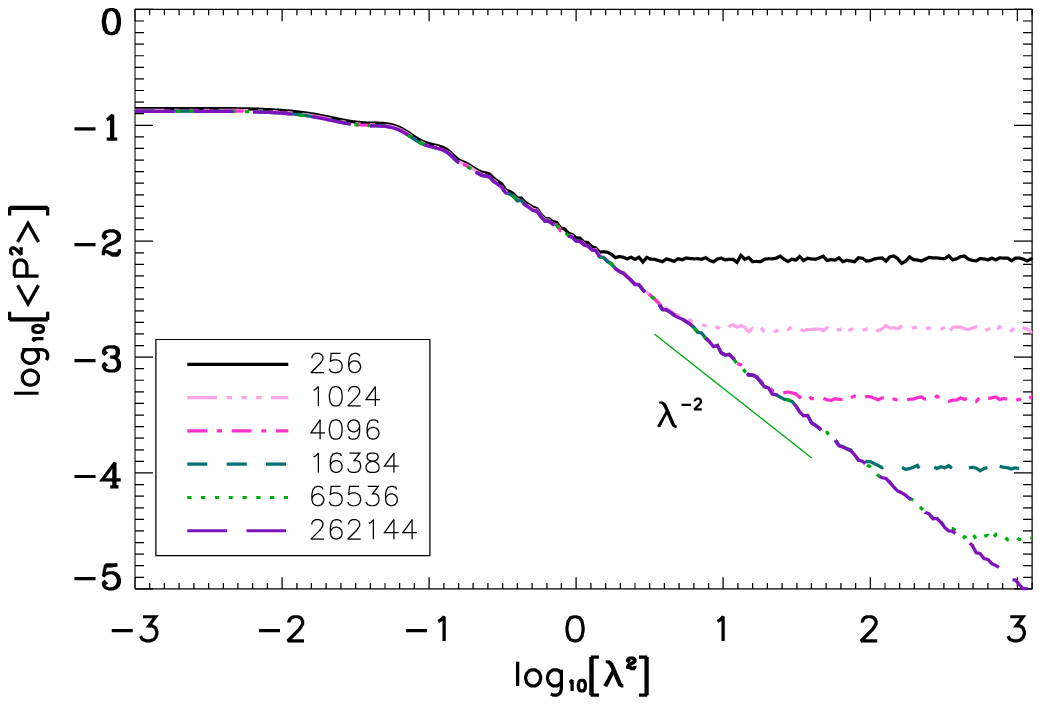}
        \caption[ ]{Same as Figure \ref{fig:var3dres} but for 1D case with different resolutions. } \label{fig:var1dres} \end{center}
\end{figure}

\begin{figure}
    \begin{center}
        \includegraphics[scale=0.7,angle=90]{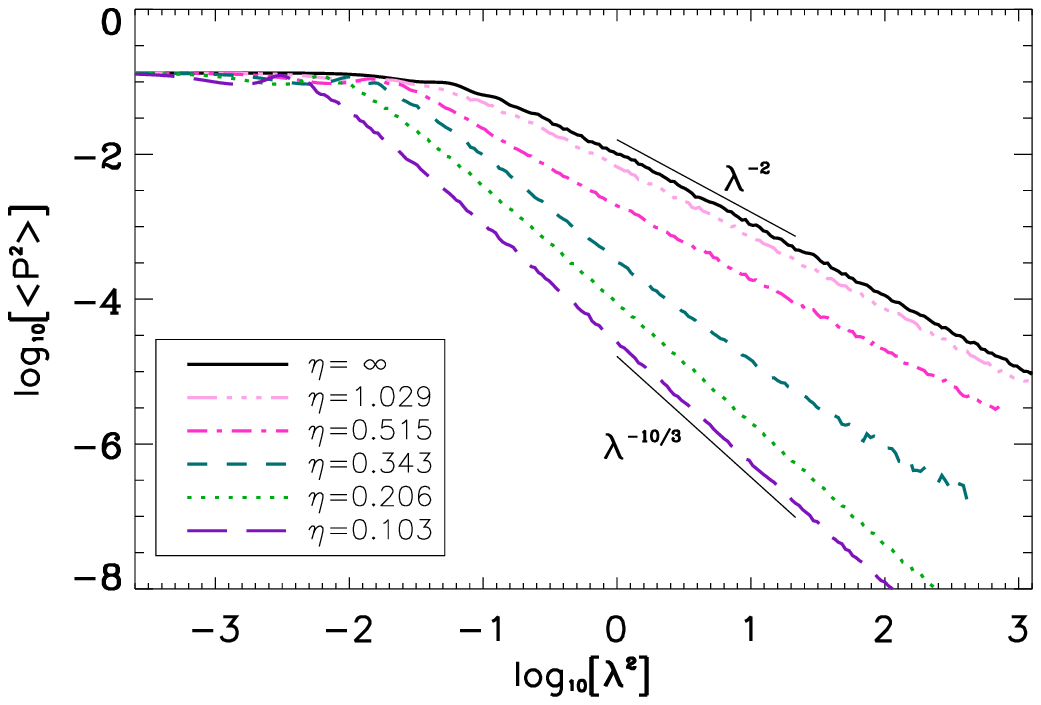}
\caption[ ]{Same as Figure \ref{fig:var3dmean} but for 1D case with different values of the mean magnetic field.
The highest resolution 1D data is used. } \label{fig:var1devol}
    \end{center}
\end{figure}

\begin{figure}
    \begin{center}
        \includegraphics[scale=0.7,angle=90]{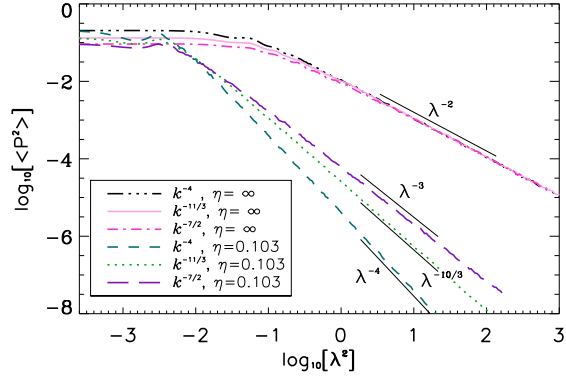}
\caption[ ]{Same as Figure \ref{fig:var3dks} but for 1D case. $\eta=\infty$ and 0.173 characterize different regions of Faraday dispersion.
The highest resolution 1D data is used.} \label{fig:var1dks}
    \end{center}
\end{figure}

\begin{figure} \begin{center}
        \includegraphics[scale=0.7,angle=90]{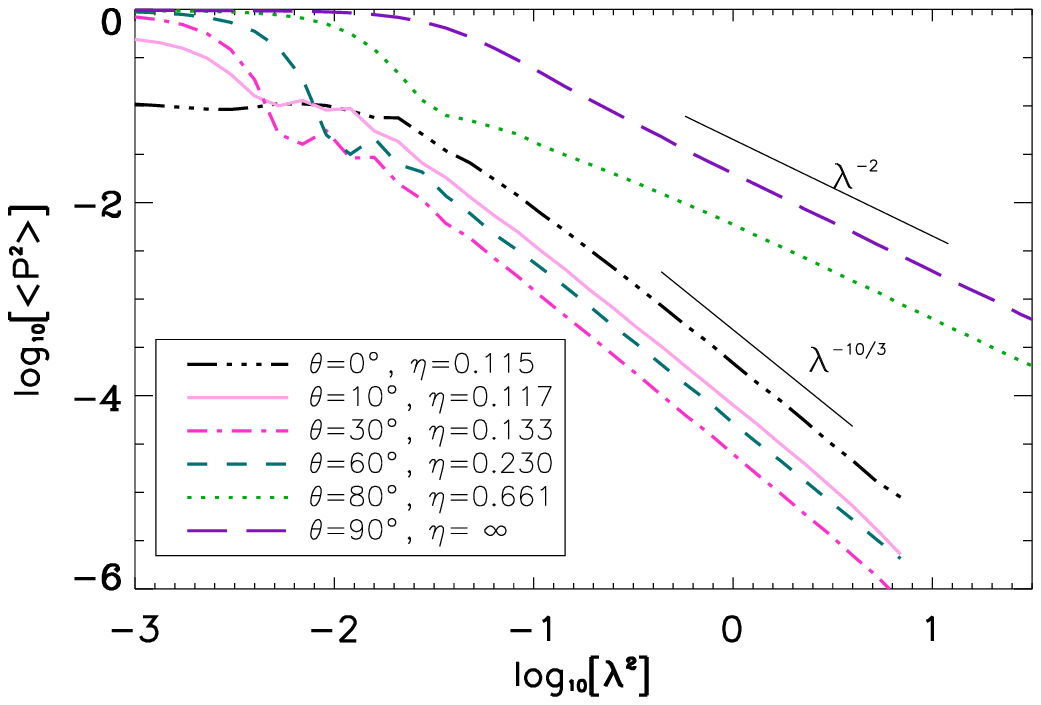}
\caption[ ]{The influence of the distribution of mean magnetic field on polarization variance in 3D case. The parameters are $\beta=11/3$ and $\left<B\right>=6$. The highest resolution 3D data cube is used.  } \label{fig:arbmf3D}
    \end{center}
\end{figure}

\begin{figure} \begin{center}
        \includegraphics[scale=0.7,angle=90]{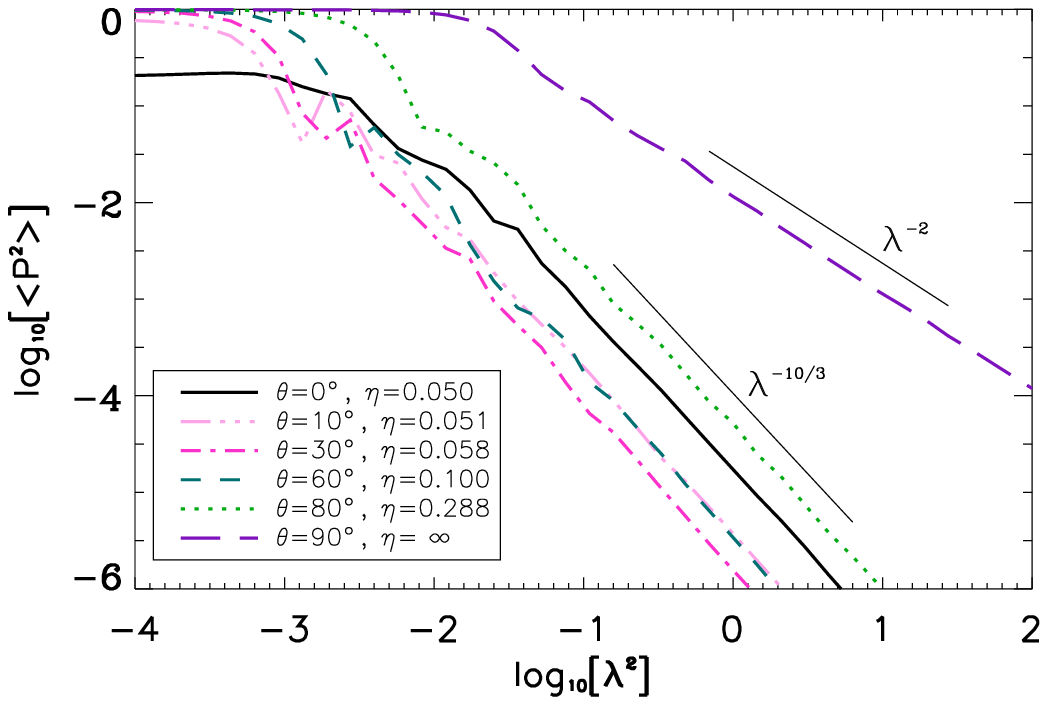}
\caption[ ]{The influence of the direction of mean magnetic field on polarization variance in 2D case. The parameters are $\beta=11/3$ and $\left<B\right>=20$. The highest resolution 2D data cube is used. } \label{fig:arbmf2D}
\end{center}
\end{figure}

\subsection{Variance of Polarization}
\label{test321D}
In this section, we test in detail the analytical PFA technique of LP16 by using 3D, 2D and 1D synthetic data `cubes', with different numerical resolutions, different values of mean magnetic fields along the $z$-axis, and various magnetic energy
spectra of $k^{-4}$, $k^{-11/3}$and $k^{-7/2}$.

\subsubsection{3D Tests}
Using the numerical technique introduced in Section \ref{NumTech}, we could in principle obtain a 3D box with any numerical resolution along each side. However, in practice, the size of the 3D grid is limited by the available memory of the computer.
The maximum numerical resolution for our 3D cubes is $4096^3$.
Note, however, that, in order to reduce memory requirement, we save data for only
$128^2$ evenly-spaced lines of sights.
Therefore, although we use the notation $128 \times 128 \times N$ for numerical resolution for 3D data, the actual numerical resolution is equivalent to $N^3$.

Figure \ref{fig:var3dres} shows polarization variance as a function of the square of wavelength $\lambda^2$. We use various numerical resolutions, zero mean magnetic field, the Komolgorov spectral index $\beta=11/3$ for magnetic spectrum, and $\gamma=2$ (see Section \ref{sect:3.2} for the definition of $\gamma$;
hereinafter, we always use this value, unless otherwise stated).
The spectral index of density spectrum is also $11/3$.
The first plateau at short wavelengths indicates the fact that the influence of Faraday dispersion on the polarization variance is negligible. It is not difficult to understand this phenomenon following Equation (\ref{PI}). If wavelength is very small, the exponential factor arising from Faraday rotation effect would approximate to 1. As a result, polarization variance would not dependent on the square of wavelengths, resulting in a plateau.
Another plateau at long wavelength region is an unphysical phenomenon due to the limited numerical resolutions. As shown in Equations (\ref{PI}) and (\ref{var2}), the spatial distance $z$ is associated with the numerical resolution\footnote{
From a numerical point of view, when integrating polarization intensity, we have to deal with the complex exponential function $e^{2i\lambda^2\Phi}$ (can be transferred into cosine and sine functions) in Equation (\ref{PI}) from Faraday rotation effect. Using general numerical integral methods, such as a rectangular integral method or trapezoidal integral method, to carry out integrals for exponential function, cosine or sine function, it is obvious that insufficient integral precision will result in rough, even spurious result. Therefore, it is necessary to provide an enough small integral step in order to obtain real result. In fact, we could also understand this behavior from another point of view. On the basis of Equation (6), we know that polarization variance is proportional to $e^{-4\lambda^4\Phi}$. When setting $\lambda=1$, the polarization variance will only be a function of spatial distance $z$, which is related to numerical resolution. Where a low numerical resolution would result in a large integral value of the polarization variance and vise verse, which corresponds to the location of plateau. If one sets a constant magnetic field and a constant density, it is easy to test that integrating $e^{-4\lambda^4\Phi}$ to the spatial distance $z$ with a low numerical resolution (equivalent to large integral step) will present a plateau in the range of large $\lambda$ region. With increasing the numerical resolution, the plateau naturally shifts towards larger $\lambda$ region.}. When the numerical resolution gradually increases, we can see that the plateau is shifting towards longer wavelength regime. If one can provide a sufficiently high numerical resolution, this unreal phenomenon would be eliminated.
Besides, with increasing pixel sizes, the slope of the curve between two plateaus
approaches $-1$ (i.e. the curve becomes compatible with $\lambda^{-2}$).
 This result agrees with the theoretical prediction in Equation (\ref{varslope2}).
 Hereinafter, we shall use 2D and 1D synthetic data to confirm this point.

Using the highest resolution 3D data cube (i.e., the one with $128 \times 128 \times 4094$ resolution in Figure \ref{fig:var3dres}),  we in Figure \ref{fig:var3dmean} study the scenario how the change of the mean magnetic field along the $z$-axis (i.e.the line of sight) affects the polarization variance. The zero mean magnetic field characterizes a stochastic RM regime where the power-law index approximates to $1.95$
(i.e. $\left< P^2 \right> \propto \lambda^{-1.95}$), and the low value of $\eta=0.040$ would correspond to a uniform RM predominant regime where the power-law index approximates to $3.33$, which is in good agreement with the theoretical prediction, i.e., $2+2m=10/3$ with $m=2/3$ for Komolgorov scaling. The transition from the stochastic RM dominant regime to the uniform RM dominant one is finished at $\eta\approx0.080$.

We explore the behavior of polarization variance by varying MHD scaling index in Figure \ref{fig:var3dks}. Three different spectral indices $\beta=4$, 11/3 and 7/2 are used to test both random and uniform RM dominant regions. As shown in Figure \ref{fig:var3dmean},  $\eta=\infty$ or low $\eta$ value can just characterize these two turbulent regimes. We hence fix $\eta=\infty$ and 0.080 for different scaling indices, respectively. It is interesting that we find that in the random RM dominant regime MHD turbulence for different types always reveals the same asymptotic $\lambda^{-1.95}$ at long wavelength region, which is close to the predicted scaling relation, $\lambda^{-2}$. Moreover, in the uniform RM dominant regime they follow a certain relation, $\lambda^{-2-2m}$, at long wavelength regime.
Specifically, $\lambda^{-3.99}$ for $\beta=4$, $\lambda^{-3.25}$ for $\beta=11/3$, and $\lambda^{-2.85}$ for $\beta=7/2$ corresponds to the theoretical prediction of LP16, $\lambda^{-2-2m}=\lambda^{-4}$ for $m=1$, $\lambda^{-2-2m}=\lambda^{-10/3}$ for $m=2/3$, and $\lambda^{-2-2m}=\lambda^{-3}$ for $m=1/2$, respectively.

We feel that we are unsatisfied to these 3D tests due to the limitation of numerical resolutions and statistical sample. Hence, we shall work with higher resolution 2D and 1D synthetic data in the following sections.

\subsubsection{2D Tests}
We first test the influence of numerical resolution on the polarization variance by using
2D data on a square gird with the same numerical resolution along each direction. The results are plotted in Figure \ref{fig:var2dres}. Similar to Figure \ref{fig:var3dres}, the overall spectral shape presents two plateaus as well. The first plateau at short wavelength regime shows a slight change in magnitude due to changes of numerical resolution.
For the highest numerical resolution shown, the power-law index at long wavelength region approximates to 1.98 (i.e. $\left< P^2 \right> \propto \lambda^{-1.98}$), which is in good agreement with the theoretical value of 2 (see also Equation (\ref{varslope2})).

We present the effect of the mean magnetic field along the line of sight
on the variance of polarization fluctuations, especially
transition from the stochastic RM dominant regime  to
uniform RM dominant one, in Figure \ref{fig:var2dmean}.
The zero mean magnetic field case characterizes the case that RM is predominantly stochastic, and the cases with strong mean magnetic fields (along the $z$-axis) correspond to the case that Faraday rotation is mainly uniform. It can be seen that the transition from the stochastic RM dominant regime to the uniform RM dominant regime occurs at $\eta\sim 0.434$.
For $\eta\lesssim0.231$, the power-law index is practically the same, i.e., $\lambda^{-3.36}$, only presenting a shift of the curve towards a shorter wavelength. It is obvious that the result matches the theoretical prediction of LP16, $\lambda^{-2-2m}=\lambda^{-10/3}$ for Komolgorov scaling slope of $\beta=11/3$.

We in Figure \ref{fig:var2dks} compare the polarization variance for different types of MHD turbulence as a function of the square of wavelength in two different regimes of Faraday dispersion, using the mean magnetic field strengths of 0 and 20 to represent these two regimes. We use three different spectral indices for magnetic spectrum: $\beta=4$, 11/3 and 7/2. We find that in the stochastic RM regime, different MHD turbulence presents the same slope approximating to 1.98 at long wavelength region. In the uniform RM regime, the power-law slopes show variations. The measured power-law slopes are $\lambda^{-3.95}$ for $\beta=4$, $\lambda^{-3.28}$ for $\beta=11/3$, and $\lambda^{-2.97}$ for $\beta=7/2$, which is in good agreement with the theoretical prediction of LP16, $\lambda^{-2-2m}=\lambda^{-4}$ for  $m=1$, $\lambda^{-2-2m}=\lambda^{-10/3}$ for $m=2/3$, and $\lambda^{-2-2m}=\lambda^{-3}$ for $m=1/2$, respectively.
These differences can allow us to estimate the spectral slope of MHD turbulence in the emitting region.

\subsubsection{1D Tests}
The numerical technique to generate 1D synthetic data is similar to that for 2D data, but we generate data only on a line along the line of sight. We find in our numerical simulations that resulting distributions of the variance of polarization present a significant fluctuation. As a result, it is difficult to measure an exact power-law index. To reduce random fluctuations, we increase the number of 1D data sets to about 2000 by altering seeds for random numbers and then average over them. Here, each random seed is equivalent to an individual line-of-sight direction and averaging procedure has a role of obtaining statistics on the whole sky plane of interest perpendicular to the line of sight.

Similar to the study of 3D and 2D cases, we first test the effect of numerical resolution on the polarization variance. The polarization variance as a function of the square of wavelength $\lambda^2$ is plotted in Figure \ref{fig:var1dres}.
We only plot the cases with zero mean magnetic field
and the different curves in the figure
correspond to different numerical resolution. The plateau at long wavelength region, which is most pronounced at the lowest numerical resolution, i.e, $256$ grid points, shifts towards longer wavelength regions with increasing numerical resolution. Henceforth, we carry out numerical studies with $2^{18}$ grid points. In this case, we obtain a constant power-law index of 1.99, which complies well with the theoretical prediction of LP16 given in Equation (\ref{varslope2}).

Figure \ref{fig:var1devol} shows transition of the variance of polarization fluctuations from zero mean magnetic field case, which corresponds to the case when RM is predominantly stochastic, to a strong mean magnetic field case, which corresponds to the case when RM is mainly due to the uniform magnetic field. As shown in this figure, the transition from the stochastic RM dominant regime to the uniform RM dominant regime occurs at $\eta\approx0.343$, and is completed at $\eta\approx0.206$. For larger $\eta$, the power-law index is practically the same, i.e., $\lambda^{-3.34}$. It is obvious that the result matches well the theoretical prediction of LP16, $\lambda^{-2-2m}=\lambda^{-10/3}$ for Komolgorov scaling slope with $\beta=11/3$.

We present the studies for different MHD spectral indices in Figure \ref{fig:var1dks}. Two representative extremal magnetic field strengths are used to characterize two different regimes of MHD turbulence , that is, $\bar{\phi}>\sigma_{\phi}$ for $\eta=0.103$ and $\bar{\phi}<\sigma_{\phi}$ for  $\eta=\infty$. In the case of $\bar{\phi}<\sigma_{\phi}$, i.e., stochastic RM dominant regime, the resulting slope does not change with the spectral index of MHD turbulence at long wavelength region, and there is only slight influence on the amplitude of variance at short wavelength region, where the RM effects is negligible.  In the opposite case of $\bar{\phi}>\sigma_{\phi}$, i.e., uniform RM dominant regime, the power-law index shows significant changes. It is very interesting that these changes still follow the relation of $\lambda^{-2-2m}$, that is, $\lambda^{-3.96}$ for $\beta=4$, $\lambda^{-3.34}$ for $\beta=11/3$, and $\lambda^{-2.96}$ for $\beta=7/2$ , which consists well with the theory prediction of LP16, $\lambda^{-2-2m}=\lambda^{-4}$ for $m=1$, $\lambda^{-2-2m}=\lambda^{-10/3}$ for $m=2/3$, and $\lambda^{-2-2m}=\lambda^{-3}$ for $m=1/2$, respectively.

\begin{figure*}[]
    \begin{center}
        \begin{tabular}{ccc}
            \hspace{-0.79cm}
            \includegraphics[angle=90,scale=0.7]{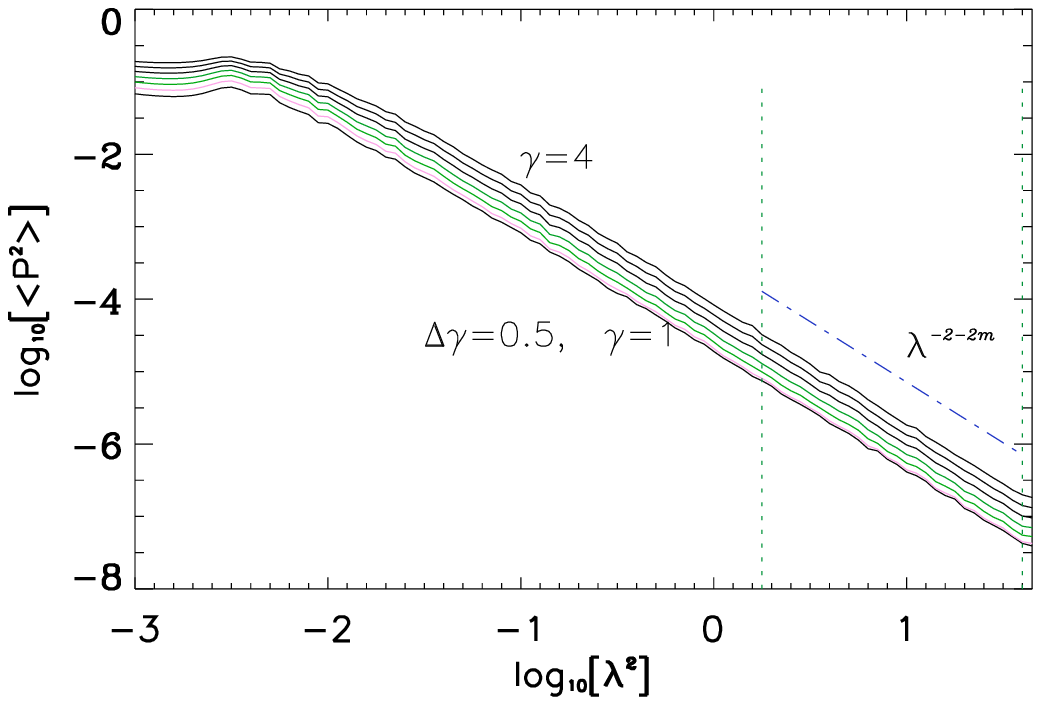}& \ \ \
            \hspace{-0.79cm}
            \includegraphics[angle=90,scale=0.7]{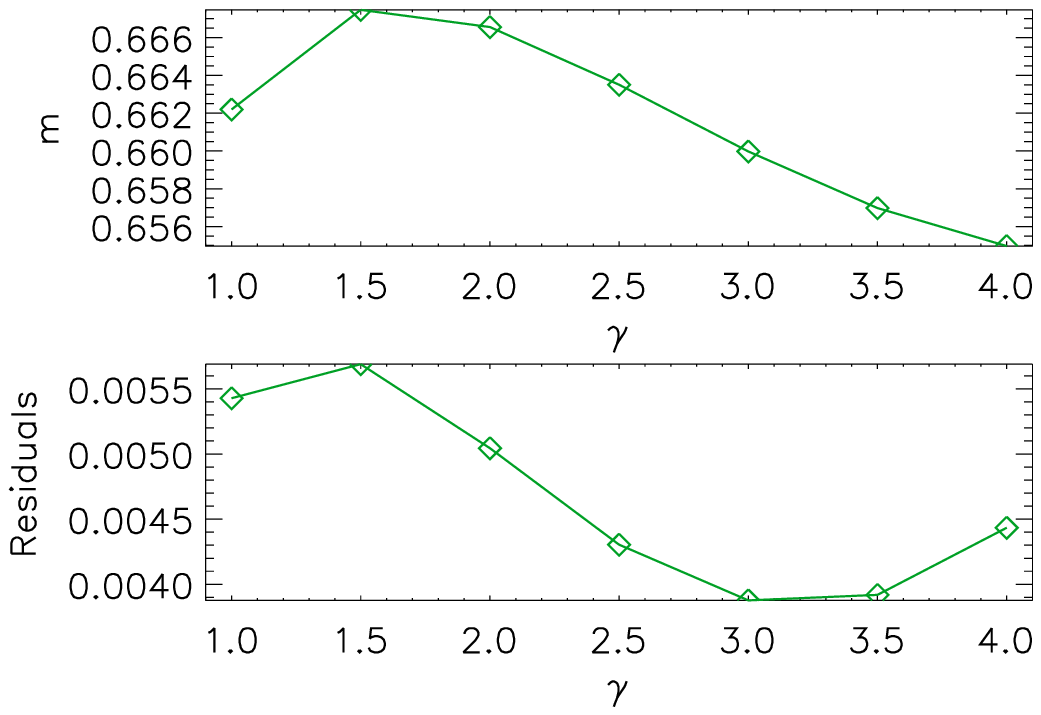}\\
        \end{tabular}
    \end{center}
    \caption{Left: polarization variance as a function of the square of wavelength, corresponding to $\gamma\in[1,4]$ in an increment of $\Delta\gamma=0.5$. Right: The spectral index of the polarization variance as a function of $\gamma$ and the resulting sum of the squared residuals of the linear fitting. The range of $\lambda^2$ used for fitting corresponds to the region between two vertical dotted lines as shown in the left panel.}  \label{figs:spec1d}
\end{figure*}

\begin{figure} \begin{center}
        \includegraphics[scale=0.7,angle=90]{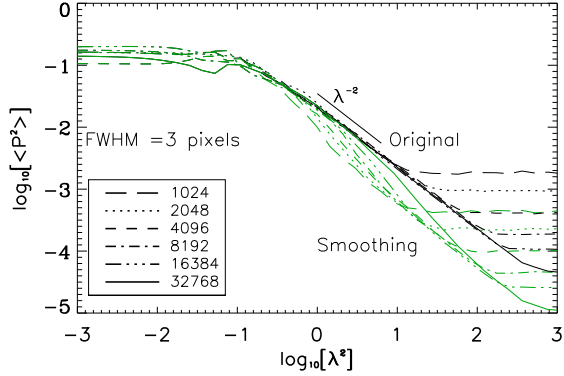}
        \caption[ ]{The influence of numerical resolutions on smoothing processes in 2D case. The parameters are $\beta=11/3$ and $\eta=\infty$. The maps corresponding to the original slopes have been smoothed to an effective Gaussian beam of FWHM=3 pixels.   } \label{fig:ressmoo}
    \end{center}
\end{figure}

\begin{figure*}[]
    \begin{center}
        \begin{tabular}{ccc}
            \hspace{-0.79cm}
            \includegraphics[angle=90,scale=0.7]{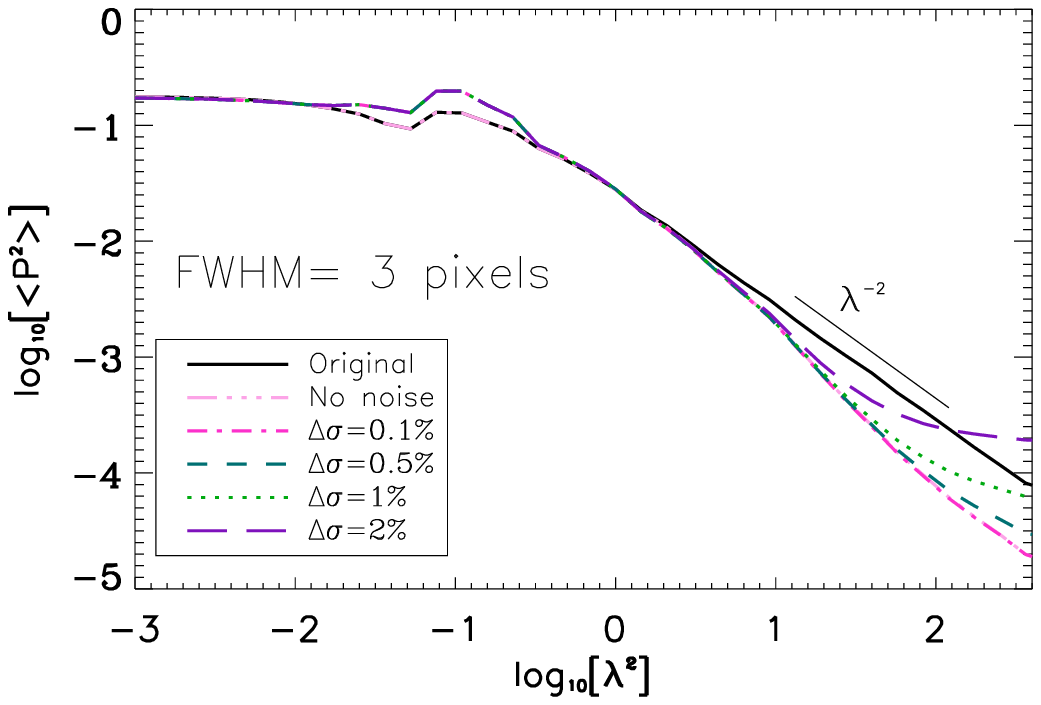}&
            \hspace{-0.79cm}
            \includegraphics[angle=90,scale=0.7]{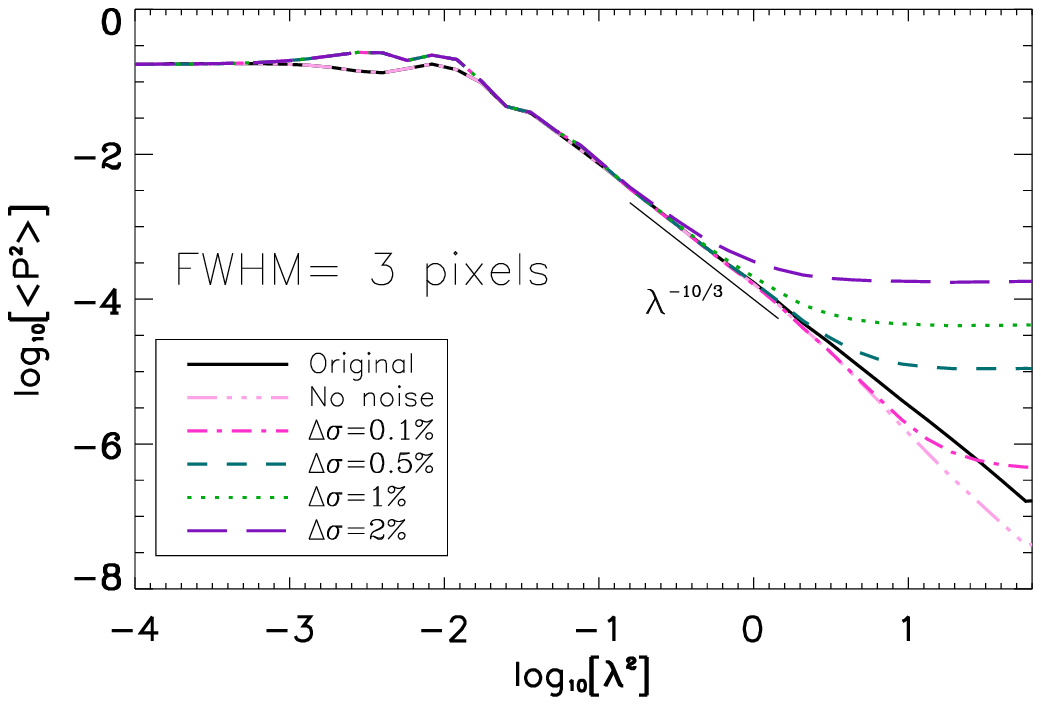}\\
        \end{tabular}
    \end{center}
    \caption{The influence of Gaussian noise on the polarization variance in the stochastic (left) and uniform (right) RM dominant regions in 2D case, for $\beta=11/3$. $\Delta\sigma$ denotes standard deviation of Gaussian noise and is a fraction of the mean synchrotron intensity. The original slope is plotted in the solid line. The maps after adding noise (corresponding slopes of polarization variance not shown for clarity) have been convolved to an effective Gaussian beam of FWHM=3 pixels at different noise levels. The highest resolution 2D data is used.}  \label{figs:noise}
\end{figure*}

\subsection{Influence of Mean Magnetic Field Distributions on Polarization Variance }
In the previous sections, we assumed that the direction of mean magnetic field is oriented along the line of sight, i.e., $z$-axis direction. However, it may present other distributions rather than only along the $z$-axis in a real astrophysical environment. In order to explore this influence on polarization variance, we consider that the existence of a mean magnetic field $\left<B\right>$ is confined in the $x$-$z$ plane for simplicity, in which the angle subtended by the field direction and the line of sight ($z$-axis) is labeled as $\theta$. Following the methods given in Section \ref{NumTech}, we first generate 3D and 2D data cubes $B_{x0}$, $B_{y0}$, $B_{z0}$, under an initial condition with zero mean magnetic field. We then project the mean field set in the $x$-$z$ plane into the $x$- and $z$-axis directions. As a result, data cubes are given as follows: $B_x=B_{x0}+\left<B\right>{\rm sin}\theta$,  $B_y=B_{y0}$, and $B_z=B_{z0}+\left<B\right>{\rm cos}\theta$.

In Figure \ref{fig:arbmf3D}, we study that the influence of the direction of mean magnetic field on the polarization variance in the 3D case. The parameters are the turbulence index $\beta=11/3$ and the mean magnetic field strength of alterable orientation $\left<B\right>=6$. Accordingly, the ratio $\eta$ would be defined as $\eta=\sigma_z/\left<B\right>{\rm cos}\theta$. As shown in this figure, two extreme scenarios, $\theta=0^\circ$ and $90^\circ$, correspond to the case when the direction of mean field is along the $z$-axis and the case when the $x$-axis, respectively. The former is similar to the case of $\eta=0.040$ presented in Figure \ref{fig:var3dmean}, but where the mean magnetic field is generated by setting corresponding initial conditions. The latter is similar to the case of $\eta=\infty$, but here there is a non-zero mean magnetic field along the $x$-axis direction. The other $\theta$ values correspond to the transition between both of them. It can be seen that they would result in different mean field level in the $z$-axis, as characterized in $\eta$. 

Figure \ref{fig:arbmf2D} shows that the influence of the direction of mean magnetic field on the polarizaton variance in the 2D case. The parameters are the turbulence index $\beta=11/3$ and the mean magnetic field of alterable orientation $\left<B\right>=20$. The related descriptions are similar to that given for Figure \ref{fig:arbmf2D}. We find in the 3D and 2D cases that for large $\eta$, the polarization variance follows usual relation $\left< P^2 \right> \propto \lambda^{-2}$, and for small $\eta$, $\left< P^2 \right> \propto \lambda^{-2-2m}=\lambda^{-10/3}$. The resulting simulations are in agreement with the theory prediction of LP16.

\subsection{Influence of Electron Spectral Index on Polarization Variance }
 We here explore how the power-law spectral slope of cosmic ray energy distribution affects the variance of polarization. Our results for variance of polarization (Section \ref{test321D}) demonstrate that the 1D result is best, so we perform this exploration based on 1D synthetic observations. The results are presented in Figure \ref{figs:spec1d} for the case of the Kolmogorov turbulence with $\beta=11/3$. The left panel plots polarization variance as a function of the square of wavelength, in which different curves correspond to $\gamma\in[1,4]$ in an increment of $\Delta\gamma=0.5$. It can be seen that the amplitude of polarization variance increases with $\gamma$, but their overall shapes seem to be similar.

The right panel presents the spectral index of the polarization variance as a function of $\gamma$ and the resulting sum of the squared residuals of the linear fitting. The range of $\lambda^2$ used for fitting corresponds to the wavelength range between two vertical dotted lines in the left panel, which coincides with the wavelength region
for obtaining the prediction of the power-law index of polarization variance, i.e., at longer wavelength region. As shown in this panel, the measured power-law indices show very small variations around the mean value of $m\approx 0.66$. As a result, variations of cosmic ray index have no influence on the power-law index of polarization variance.

\subsection{Observational Influence on Polarization Variance }
The technique for adding observational effects, i.e., effects of both finite angular resolution and noise, introduced in Section \ref{NumTech}, has a universality for relevant studies. For instance, it can be used to study the influence of observations on the structure function and power spectrum. We here focus on the study of observational influences on the variance of polarization.

Using a 3D synthetic data with 512 pixels (i.e., grid points) along each side, we study the influence of both angular resolution and noise on the power-law slope of the polarization variance. At each wavelength, the maps of Stokes $I$, $Q$ and $U$ are produced, which is called the idealized (original) images.
Provided that one assumes that 1 pixel corresponds to 0.05 pc and the distance of the emission region is 2 kpc away, one would obtain an angular distance of 5 arc seconds for 1 pixel. With a Gaussian kernel, the original maps can be convolved to an expected pixel that corresponds to the standard deviation of radio telescope beam. Furthermore, the original images are added with a Gaussian noise with standard deviation equal to a fraction of the mean synchrotron intensity, and then new images are smoothed to FWHM=3 pixels. Using these images, we calculate observed power-law slopes of the polarization variance. However, we find that the resulting slopes are deviated from the original forms, $\lambda^{-2}$ and $\lambda^{-2-2m}$. The reason is that the 3D box with 512 pixels along each side does not have enough resolution. Therefore, we switch to 2D data by significantly enhancing numerical resolutions.

\begin{figure*}[]
    \begin{center}
        \begin{tabular}{ccc}
            \hspace{-0.79cm}
            \includegraphics[angle=90,scale=0.7]{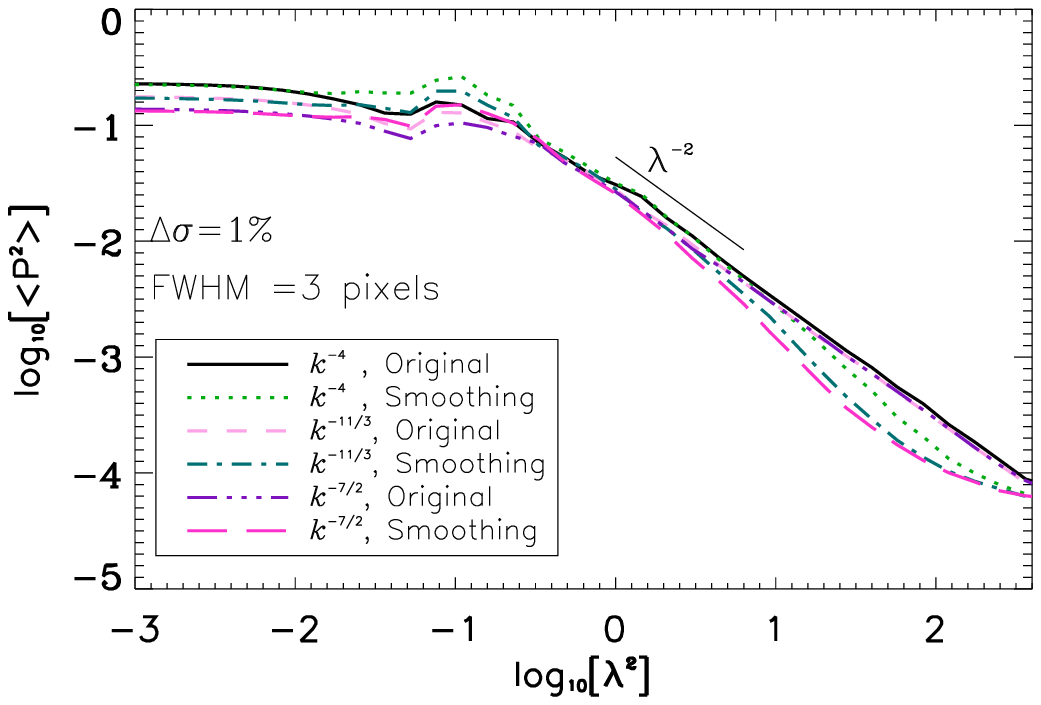}&
            \hspace{-0.79cm}
            \includegraphics[angle=90,scale=0.7]{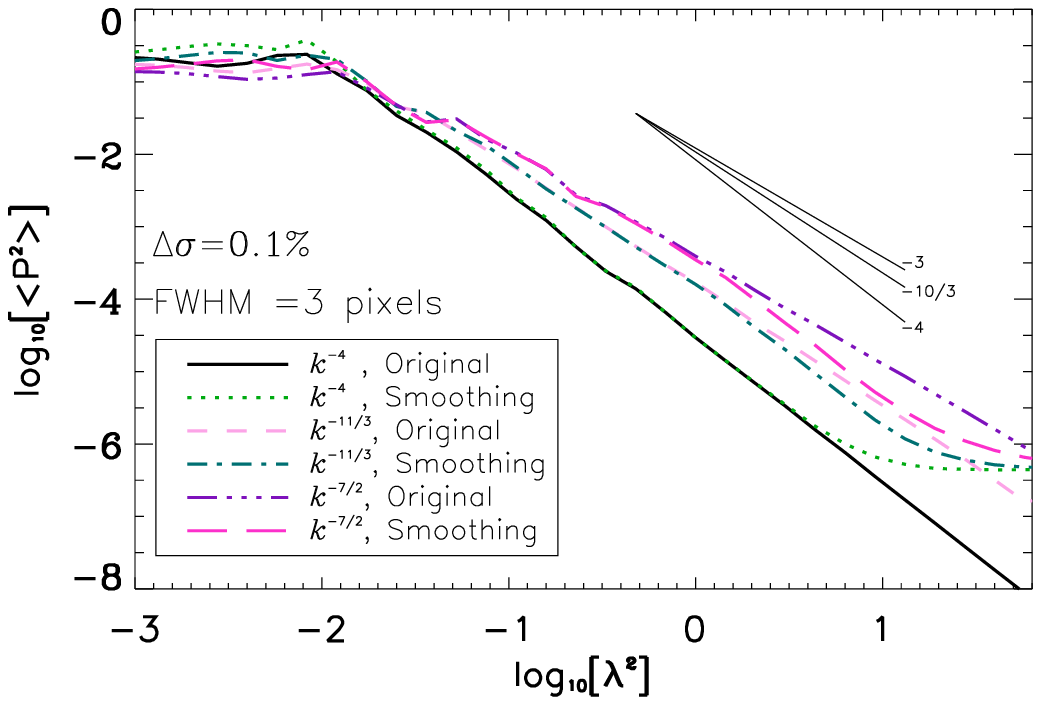}\\
        \end{tabular}
    \end{center}
\caption{The resulting slopes of the polarization variance for different slope indices of MHD turbulence in 2D case, corresponding to the stochastic (left) and uniform (right) RM dominant regions. Both angular resolution and Gaussian noise are fixed. The highest resolution 2D data is used.}  \label{figs:final}
\end{figure*}

The technique for studying observational effects in 2D case is similar to that in 3D. Only difference is that the observed maps are one-dimensional. The influence of numerical resolutions on smoothing processes is presented in Figure \ref{fig:ressmoo}. The original maps have been smoothed with an effective Gaussian beam of FWHM=3 pixels. As shown in the figure, the curves for the smoothed maps get closer to those of original maps as  numerical resolution increases. We hence use the data with
maximum numerical resolution we can provide to carry out the following research.

The influence of Gaussian noise levels on $\left< P^2 (\lambda^2) \right>$ 
in the stochastic and uniform RM dominant regimes is plotted in Figure \ref{figs:noise}, for a Komolgorov magnetic spectrum with $\beta=11/3$. The standard deviation of Gaussian noise is parameterized by $\Delta\sigma$ that is a fraction of the mean synchrotron intensity. The maps after adding Gaussian noise have been convolved to an effective Gaussian beam of FWHM=3 pixels at different noise levels. We find that different noise levels indeed
produce different results. The Gaussian noise makes the polarization variance curved upwards at long wavelength region. However, as seen in Figure \ref{fig:ressmoo}, slopes corresponding to smoothed maps would deviate downwards from the original slope. As a result, from the narrow wavelength point of view, an adding noise could make spectrum more close to the original slope. Thus, it needs a coverage range of wavelengths in observations in order to recover real properties of MHD turbulence.

The resulting $\left< P^2 (\lambda^2) \right>$ 
with both fixed angular resolution and fixed noise level are given in Figure \ref{figs:final}, corresponding to stochastic (left panel) and uniform (right panel) RM dominant regimes. It is obvious that steep magnetic spectrum is more easy to recover the universal scaling $\lambda^{-2}$, and $\lambda^{-2-2m}$ at the same angular resolution. Besides, to recover $\lambda^{-2-2m}$ slope, it needs higher signal to noise ratio when comparing the left and right panels. Using the method given in Section \ref{NumTech}, we deduce what the signal to noise ratio at individual wavelength is. For instance, the signal to noise ratios at $\lambda^2\sim40$, corresponding to the noise levels of $\Delta\sigma=0.1\%$, $0.5\%$, $1\%$ and $2\%$, are 158, 32, 16 and 8, respectively. The noise has a small influence on the polarization variance at slightly short wavelength region which is an ideal region of wavelengths for recovering the properties of magnetic turbulence. The uniform RM dominant regime is important for us to reveal the properties of magnetic turbulence from the polarization variance, because the $\lambda^{-2-2m}$ scaling is associated with different types of MHD turbulence by the spectral index $m$.

\subsection{Depolarization and Variance of Polarization Derivative }
LP16 provides additional methods for extracting statistical information
on polarization fluctuations. As examples, we in this section present studies of the degree of polarization, and the variance of polarization derivative with respect to the square of wavelength ($\left< (dP/d\lambda^2)^2 \right>$), based on 1D synthetic data.

\begin{figure} \begin{center}
        \includegraphics[scale=0.7,angle=90]{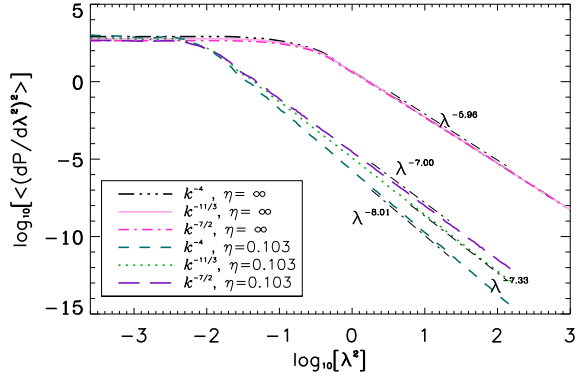}
\caption[ ]{The variance of polarization derivative with regard to $\lambda^2$ for different MHD scaling indices, as a function of the square of wavelength $\lambda^2$ in 1D case. $\eta=\infty$ and 0.103 correspond to different regions of Faraday dispersion.
The highest resolution 1D data is used.} \label{fig:dervar}
    \end{center}
\end{figure}

Figure \ref{fig:dervar} shows the variance of polarization derivative with respect to $\lambda^2$ for different magnetic spectral indices $\beta=4$, 11/3 and 7/2. We use two extremal values of mean magnetic field along the line of sight, which characterize different regimes of Faraday dispersion. In the case of $\eta=\infty$, the resulting scaling follows $\lambda^{-5.96}$ at long wavelength region. In the case of $\eta=0.103$, the power-law indices are $\lambda^{-8.00}$ for $\beta=4$, $\lambda^{-7.33}$ for  $\beta=11/3$, and $\lambda^{-7.00}$ for $\beta=7/2$, respectively. Although there is not a theoretical prediction, we believe that this method can surely recover the scaling properties of MHD turbulence.

\begin{figure} \begin{center}
        \includegraphics[scale=0.7,angle=90]{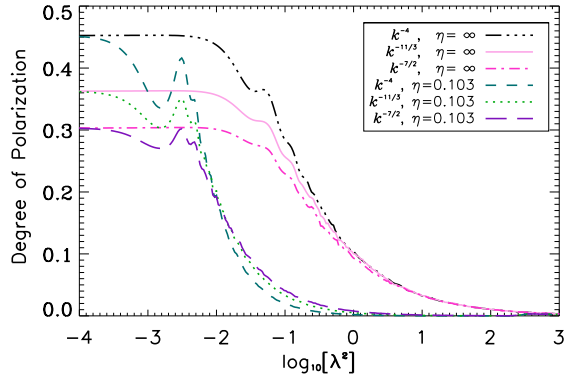}
        \caption[ ]{Degree of polarization of linearly polarized synchrotron emission as a function of the square of wavelength $\lambda^2$ in 1D case. The potential scaling indices of MHD turbulence are set as $\beta=4$, 11/3 and 7/2. $\eta=\infty$ and 0.103 correspond to different regions of Faraday dispersion.
        The highest resolution 1D data is used.} \label{fig:dofp}
    \end{center}
\end{figure}

We in Figure \ref{fig:dofp} present the degree of polarization of linearly polarized synchrotron emission as a function of the square of wavelength $\lambda^2$. As shown in the figure, the polarization for steep magnetic spectral index is stronger than that of shallow spectral index. Correspondingly, the depolarization of Faraday dispersion is also stronger in steep slope than in shallow one. The increase in the mean magnetic field along the line of sight results in the shift of effective Faraday dispersion towards much shorter wavelength regime. It is easy to understand this point from the relation $\chi\propto\lambda^2(\bar{B_z}+B_{z})$, in which $\chi$ is the Faraday
rotation angle of the polarized direction. When $\chi$ is invariant, the increase in the magnetic field shall result in a decrease in $\lambda^2$. The random fluctuations in the degree of polarization caused by Faraday dispersion become significantly larger in large mean field than zero mean field, in particular, at $\lambda^2\approx0.003$.

\section{Discussion}

We have proven that using the dispersion of synchrotron polarization one can obtain the statistics of the magnetic field and the underlying statistics of Faraday rotation. In the case of negligible Faraday rotation, the statistics of synchrotron polarization provide spectral properties of the perpendicular component of the magnetic fields (LP12). But as described in LP16, due to the presence of an extra imaginary part of the polarization intensity, resulting statistical studies can deliver more valuable information. For example, the imaginary part of the correlation functions of polarization intensity could provide information about the 3D direction of the mean magnetic field, which is a very intriguing result in LP16. In the case when Faraday rotation is important, one can obtain statistical information about fluctuations of the magnetic field as well as thermal electron density. One can also sample different emission regions by varying wavelengths, which is the most obvious advantage of Faraday dispersion and provide the valuable information about turbulence distribution along the line of sight.

In the present study, we are focused on testing the PFA technique of LP16 by using 3D, 2D and 1D synthetic data `cubes'. Our study probes turbulence along a line of sight direction. A setup with zero mean magnetic field would characterize the stochastic RM dominant region, in which we confirmed that the variance of polarization intensity follows the universal $\lambda^{-2}$. Thus, by itself, such a study cannot provide the properties of turbulence. As suggested in LP16 a derivative of polarization variance with regard to the square of wavelengths allows to recover the statistical information about the perpendicular component of magnetic fields, but the procedure of taking a derivative enhances noise which requires a more sophisticated regularization procedures. Therefore we have not tried it in this paper.  With increase of mean magnetic fields, the stochastic Faraday rotation dominant region would transform into the uniform RM dominant region, in which we confirmed that the variance of polarization intensity follows $\lambda^{-2-2m}$. This region provides a way for us to recover the spectral properties of magnetic turbulence.

It is noticed that when we carry out simulations based on different dimensions, i.e., 3D, 2D and 1D, the typical value of the ratio $\eta$ corresponding to the shift from the turbulent dominated region to the mean field dominated region is not the same. For 3D synthetic data, we have considered a sophisticated divergence-free condition of the synthetic magnetic field. In order to significantly enhance numerical resolutions, we have used a fractal cube for 2D and 1D data. A slight difference to generate synthetic data cube with different dimensions would result in a difference in turbulent magnetic field. Besides, the structure of the integrals to calculate the polarization variance depends on the dimension of the space of integration. Specifically, the polarization variance depends on the synthetic magnetic fields, $B_x$, $B_y$ and $B_z$. These turbulent magnetic field components are further related to spatial coordinates $xyz$ for 3D case, $xz$ for 2D case, and $z$ for 1D case.

We have confirmed that the variance of polarization derivative with regard to $\lambda^2$ can also reveal the spectral properties of magnetic turbulence and this motivates a further theoretical studies. Similar to the traditional Faraday RM synthesis method mentioned in \S 1, other features related to Faraday rotation, such as the degree of polarization and rotation angle of polarization plane, can also be obtained in the framework of the new techniques. As an example, we explore the depolarization effect of Faraday rotation.

The goal of our paper is to pave the way for the PFA technique suggested in LP16 for applying to observations. As we know, the spectral recovery of MHD turbulence would be affected by both the noise and the resolution of telescopes. Hence, this study also explores the influence of these effects on the PFA. To test the effects of the resolution, we found that high numerical resolutions  is required to provide a sufficiently long inertial range that is necessary for recovering the turbulence spectrum. This is not a problem, however,  for actual astrophysical observations, as they do not have the issue of the shot noise that plague our numerical studies. We note that the problem of shot noise related to the discrete numerical data was a serious issue for testing VCA and VCS techniques (see \citealt{Laz01,ChepL09}). With the numerical samples with the highest resolution we successfully demonstrated the possibility of recovering turbulence data. This proves the feasibility of the technique.

Though the fact that the index of cosmic rays is not a constant, the present study confirms the  LP12 and LP16 conclusion that the change of electron index does not distort their the value of the power slope of the fluctuations measured by the PFA. Therefore, we numerically proved that the variations of cosmic ray properties does not prevent one to obtianing the spectral properties from synchrotron polarization fluctuations.

The importance of this work is obvious from yet another perspective.
It is advantageous to use synergies of different techniques to study MHD turbulence. The VCA and VCS techniques mentioned in Section 1 can be used to obtain statistics of velocity fluctuations.
In addition, we should stress that a spectrum is not the only property of turbulence. For instance, recently, some new techniques, based on the measurements of the kurtosis, skewness and genus of the gradients of the polarized synchrotron emission were applied to understand the Mach number of turbulence (\citealt{Gaensler11,Burkhart12}). Other new techniques of obtaining the statistics of density are recently developed by the analysis of moments of the density probabilities (\citealt{Kowal07,Burkhart09,Burkhart10}), Tsallis statistics (\citealt{EL10,Toff11}), bispectra and genus (\citealt{L99,Chep08,Burkhart09}). Using the new PFA technique, one can obtain the statistics of magnetic fields and Faraday rotation. This information is complementary for the turbulence studies.

Our work is intended to open an avenue towards analyzing the large amount of exiting and upcoming radio data cubes from LOFAR and SKA, with the aim of applying the new technique to the study of the Milky Way and Galaxies.

\section{Summary}
In this paper, we have used synthetic observations to test the analytical predictions of the new statistical techniques suggested in LP16. We found that numerical results are in good agreement with the LP16's study. Based on 3D, 2D and 1D synthetic observations, numerical results we have obtained are briefly summarized as follows.

1. The studies of the polarization dispersion of synchrotron radiation demonstrated that for the region dominated by stochastic RM fluctuations, corresponding to the ratio $\eta\gg1$, the variance of polarization gives the universal spectrum $\lambda^{-2}$. In a region with the dominant RM effect arising from the regular magnetic field ($\eta\lesssim0.2$), the variance of polarization follows $\lambda^{-2-2m}$.  Thus the variance reflects statistics of the magnetic field component perpendicular to the line of sight.

2. The studies of the dispersion of polarization derivative with regard to $\lambda^2$ shows a power-law relation, which should reflect the fluctuation statistics of Faraday rotation.

3. The spectral index of relativistic electrons does not change the slope of the PFA measure, and thus does not prevent us from extracting spectral properties of magnetic turbulence.

4. The PFA technique can be practically used, as we showed that the effects of angular resolution and inevitable observational noise do not present an obstacle for recovering the underlying spectra of turbulence.

\ \

\acknowledgments  We highly appreciate the referee for his/her valuable comments and suggestions that significantly improve our work. J.F.Z. thanks support from the National Natural Science Foundation of China (grant Nos. 11233006 and 11363003). A.L. acknowledges the NSF grant AST 1212096 and the Center for Magnetic Self Organization (CMSO). H.L. and J.C. acknowledge supports by the National R \& D Program through the National Research Foundation of Korea, funded by the Ministry of Education (NRF-2013R1A1A2064475).

\end{document}